\documentclass[12pt]{article}          
\usepackage{amsmath}             
\usepackage{graphicx}            \usepackage{amssymb}
\usepackage{appendix}
\usepackage{slashed}
\usepackage{empheq}
 \font\tenrm=cmr10 

\usepackage{geometry}

\title{``Formal" vs. ``Empirical" Approaches to Quantum-Classical Reduction}  
\author{}                                       
\date{}

\begin{document}             
\maketitle

\begin{abstract}             
\baselineskip=10pt \tenrm
I distinguish two types of reduction within the context of quantum-classical relations, which I designate ``formal" and ``empirical." Formal reduction holds or fails to hold solely by virtue of the mathematical relationship between two theories; it is therefore a two-place, \textit{a priori} relation between theories. Empirical reduction requires one theory to encompass the range of physical behaviors that are well-modeled in another theory; in a certain sense, it is a three-place, \textit{a posteriori} relation connecting the theories and the domain of physical reality that both serve to describe. Focusing on the relationship between classical and quantum mechanics, I argue that while certain formal results concerning singular $\hbar \rightarrow 0$ limits have been taken to preclude the possibility of reduction between these theories, such results at most block reduction in the formal sense. Little if any reason has been given for thinking that they block reduction in the empirical sense. I then briefly outline a strategy for empirical reduction that is suggested by work on decoherence theory, arguing that this sort of account remains a fully viable route to the empirical reduction of classical to quantum mechanics and is unaffected by such singular limits. 
\end{abstract}

\section{Introduction}

Work on quantum-classical relations encompasses a vast and disparate range of results, from analyses of $\hbar \rightarrow 0$ limits and various quantization procedures to Ehrenfest's Theorem, environmental decoherence, decoherent and consistent histories, the measurement problem, interpretation-specific accounts of classicality and much else.
As Landsman has emphasized, our understanding of the relationships among these different areas is still in its infancy \cite{landsmanClassicalQuantum}. 
For the discussion that follows, it will be useful to distinguish two broad and occasionally overlapping categories of analysis in the study of quantum-classical relations, which serve to address two distinct but related sorts of question concerning the relationship between quantum and classical theories:

\begin{enumerate}
\item \textbf{``Formal" -} What is the nature of the relationship between the mathematical formalisms of classical and quantum mechanics? Classical mechanics is formulated in a mathematical arena of symplectic manifolds, canonical transformations, Poisson brackets, action principles and the like, while quantum mechanics is formulated in a realm of Hilbert spaces, unitary transformations, commutators, path integrals, $C^{*}$ algebras, PVM's, POVM's and related structures. What connections, analogies and correspondences can we identify between these two mathematical frameworks? 
\item \textbf{``Empirical" -} Under what circumstances are the behaviors described by classical and quantum models manifested \textit{in real physical systems}? How does the set of real-world cases that are well-described by some classical model relate to the set of real-world cases that are well-described by some quantum model?  
\end{enumerate}

\noindent More narrowly, and more to my central point here, it is important to distinguish ``formal" and ``empirical" approaches to \textit{reduction} between classical and quantum mechanics, where reduction broadly speaking is taken to require that one theory encompass the other in some sense. In the contexts of both formal and empirical reduction, I will adopt the convention here of referring to the less encompassing theory as the ``reduced" theory and the more encompassing theory as the ``reducing" theory, so that the former is said to ``reduce to" the latter (the opposite convention is often adopted in the physics literature). Formal and empirical reduction are distinguished as follows:

\begin{enumerate}
\item \textbf{Formal Reduction} requires the reduced theory (in this case, classical mechanics) to be in some sense a special or limiting case of the reducing theory (in this case, quantum mechanics). The question of whether one theory reduces to the other is a wholly mathematical, \textit{a priori} question to be resolved entirely through mathematical analysis of the two theories. Once the mathematical frameworks of the theories have been specified, no further empirical input is required to assess whether one reduces to the other. Formal reduction in thus a \textit{two}-place relation between theories. 
\item \textbf{Empirical Reduction}
requires that every circumstance under which the behavior of a real physical system can be modeled in the reduced theory (in this case, classical mechanics) is also one in which that same behavior can be modeled at least as precisely in the reducing theory (in this case, quantum mechanics). That is, empirical reduction requires that the reducing theory wholly subsume the physical domain of applicability of the reduced theory, but does not necessarily require the reduced theory's formal mathematical structure to be subsumed wholesale as a special or limiting case of the reducing theory's formal structure. Unlike formal reduction, the question of whether empirical reduction holds between two theories cannot, in general, be determined solely by analysis of the theories themselves. Once the theories have been specified, reduction is still partially an empirical matter, for demonstrating that one theory has subsumed the domain of applicability of the other requires empirical knowledge of the sets of circumstances under which the theories succeed at describing the behavior of real systems. Empirical reduction is therefore in some sense a \textit{three}-place relation connecting the two theories and the domain of physical reality that the theories serve to describe. 
\end{enumerate}

\noindent Thus, formal reduction requires subsumption at the level of the mathematical formalisms of the two theories, while empirical reduction requires subsumption at the level of the real physical behavior that can be accurately (if approximately) modeled in the theories. It is conceivable that one could successfully effect an empirical reduction \textit{by means} of a formal reduction: if the reduced theory is a special case of the reducing theory, then any physical behavior that is well-modeled in the reduced theory can, \textit{a fortiori}, also be modeled in the reducing theory. On the other hand, reduction in the empirical sense does not necessarily require reduction in the formal sense. It is possible that one could show one theory to subsume the domain of another without showing that the mathematical formalism of the latter constitutes a special or limiting case of the former. In particular, it is possible for the mathematical structures of two theories to dovetail approximately over some restricted domain (namely, the domain of physical reality well-modeled by the reduced theory) without either theory being a special or limiting case of the other.

The subject of reduction often arises in the context of discussions about the ``imperialism" of physics - that is, the notion that theories in physics grow ever more universal and precise in their depictions of physical reality, and that each successive theory wholly encompasses the domain of physical behavior that can be successfully modeled by its predecessor. It is clear that reduction in the empirical sense suffices for this purpose. By contrast, in requiring one theory to be a special or limiting case of another, formal reduction demands much more than is necessary to uphold the conventional wisdom that our theories grow ever more precise and universal in their physical scope. A weaker condition, which demands approximate agreement between theories only in the restricted domain where the reduced theory is successful, is sufficient for this subsumption of physical domains to occur. Unlike formal reduction, reduction in this weaker (but still highly nontrivial) empirical sense does not require the reducing theory to recover features of the reduced theory in cases where the reduced theory does not describe the behavior of any real physical system. 

Like formal reduction, empirical reduction may rest on direct mathematical correspondences between the theories that serve to ensure approximate agreement between the theories over the appropriate domain. However, the question of whether any approximate dovetailing between theories suffices for one theory to encompass the domain of success of the other can only be answered by considering further empirical input that delineates the set of cases in which the reduced theory succeeds at tracking the behavior of real physical systems, as well as the margins of error and timescales within which it does so. Therefore, knowledge only of the theories themselves is not generally sufficient to determine whether one reduces to the other in the sense that is relevant to the conventional imperialist wisdom about the progress of physics.

Here, I focus on distinguishing formal and empirical approaches to reduction in the context of quantum-classical relations. 
In Section \ref{Formal}, I consider one attempt at quantum-classical reduction that is considered in the work of Batterman and Berry, which is based on a formal recipe due to Maslov for constructing wave functions out of classical Lagrangian surfaces \cite{BattermanDD}, \cite{BerryAsymptotics}, \cite{berry1983semiclassical}. In their analysis, Batterman and Berry draw the general conclusion that reduction between classical and quantum mechanics fails because of difficulties that Maslov's construction encounters with singular $\hbar \rightarrow 0$ limits. However, their use of the term ``reduction" in this context is ambiguous, and it is unclear whether their arguments are meant to rule out reduction in a formal sense, an empirical sense, or both. Here, I argue that while their arguments may cast some doubt on the possibility of effecting reduction in a formal sense, they offer no reason for thinking that singular $\hbar \rightarrow 0$ limits should block reduction in the empirical sense. In Section \ref{Physical}, I outline an account of empirical reduction between classical and quantum mechanics that is drawn from the literature on decoherence and that is spelled out in greater detail in \cite{RosalerIN}. I argue that this analysis provides a viable account of the empirical reduction of classical to quantum mechanics and that there is little reason to think that it is impeded by singular $\hbar \rightarrow 0$ limits. In Section \ref{Conclusion}, the Conclusion, I argue that there are important lessons to be drawn from juxtaposing this empirical, decoherence-based picture of quantum-classical reduction with the more formal approach considered by Berry and Batterman: namely, that the limitations of certain formal correspondences between classical and quantum mechanics do not necessarily block reduction between these theories in the empirical sense, and that it is important generally to take care to distinguish between formal and empirical reduction when asserting the success or failure of reduction between a given pair of physical theories.

\section{``Formal" Approaches to Quantum/Classical Reduction} \label{Formal}

In a standard graduate-level textbook on quantum mechanics, one is likely to encounter some remark to the effect that classical mechanics can be recovered from quantum mechanics in the limit $\hbar \rightarrow 0$. Take Sakurai's popular monograph, \textit{Modern Quantum Mechanics}, as an example. Sakurai notes that on inserting the polar decomposition $\psi(x,t) = R(x,t) e^{i \frac{S(x,t)}{\hbar}}$ of the wavefunction into Schrodinger's equation, one arrives at the result that the phase $S(x,t)$ satisfies the classical Hamilton-Jacobi equation in the limit $\hbar \rightarrow 0$, and concludes that ``not surprisingly, in the $\hbar \rightarrow 0$ limit, classical mechanics is contained in Schrodinger's wave mechanics" \cite{sakurai1995modern}. It is not clear whether such a remark is intended simply to highlight an interesting formal correspondence between the mathematical frameworks of quantum and classical mechanics or is intended to have some deeper physical signifcance. In its claim that classical mechanics has been shown to be ``contained in" Schrodinger's wave mechanics, such a remark does seem tentatively to suggest that we can understand by virtue of this result why classical mechanics succeeds where it does given that quantum mechanics is the more ``correct" of the two theories. 

However, a little thought shows that at most, this formal result is a small part of a much more complicated story about how quantum mechanics encompasses the domain of classical mechanics, if indeed it does. First, the physical significance of this result is left obscure since it is not made clear in Sakurai's presentation what taking the limit $\hbar \rightarrow 0$ corresponds to physically given that $\hbar$ is a constant for all real systems. Furthermore, this result gives no hint of the manner in which quantum superpositions are supposed to give way to the determinate states that characterize realistic classical behavior, or of the manner in which quantum interference effects come to be suppressed in these systems. Both of these points must be addressed in any realistic quantum mechanical account of classical behavior. 


A more sophisticated examination of formal approaches to the quantum-classical correspondence, including explorations of the $\hbar \rightarrow 0$ limit, the $N \rightarrow \infty$ limit and various formal quantization procedures, as well as a discussion of decoherence, is given in Landsman's  \cite{landsmanClassicalQuantum}. He concludes that none of these ideas in isolation is capable of explaining the classical world, but suggests that ``there is some hope that by combining  all three of them, one might do so in the future." Landsman offers an extensive survey of mathematical correspondences between quantum and classical theories that have been uncovered in the literature, but is careful to note that he does not intend to address the issue of inter-theoretic reduction between classical and quantum mechanics, which will be my main focus here.  

One prominent recent analysis of reduction between classical and quantum mechanics has been advanced in the work of Batterman and Berry, both of whom argue that the singular nature of certain $\hbar \rightarrow 0$ limits precludes reduction between these two theories. Here, I argue that the sense of reduction that they adopt is ambiguous as to whether a formal or empirical sense of reduction is intended, and that once we take care to make this distinction, it becomes clear that while their arguments may pose difficulties for reduction in the formal sense, they pose little if any obstacle to reduction in the empirical sense. I begin by reviewing Berry's general view on the relationship between limits and inter-theoretic reduction in physics and then go on to discuss Batterman's closely related argument that the singular nature of the $\hbar \rightarrow 0$ limit blocks reduction of classical to quantum mechanics. 

\subsection{Berry's Analysis of Reduction in Physics}

Concerning the general methodology of reduction in physics, Berry writes, 

\begin{quote}
To begin, realise that theories in physics are mathematical; they are formal systems, embodied in equations. Therefore we can expect questions of reduction to be questions of mathematics: how are the equations, or solutions of equations of one theory, related to those of another? The less general theory must appear as a particular case of the encompassing one, as some dimensionless parameter - call it $\delta$ - takes a particular limiting value. A general way of writing this scheme is
\begin{quote}
encompassing theory $\rightarrow$ less general theory as $\delta \rightarrow 0$
\end{quote}
Thus reduction must involve the study of limits, that is asymptotics. The crucial question will be: what is the nature of the limit $\delta \rightarrow 0$? We shall see that very often reduction is obstructed by the fact that the limit is \textit{highly singular}. Moreover, the type of singularity is important, and the singularities are not only directly connected to the existence of emergent phenomena but underlie some of the most difficult and intensively-studied problems in physics today \cite{BerryAsymptotics}. 
\end{quote}

\noindent It seems more natural here to interpret Berry as adopting a formal sense of reduction.
That is, Berry seems to regard reduction as a purely mathematical, \textit{a priori} relation between the mathematical frameworks of the theories. 
However, it is possible that Berry also intends to include all forms of reduction, including empirical reduction, in his assertion that singular limits block reduction. In the case of classical and quantum mechanics, this would imply that because certain limits as $\hbar \rightarrow 0$ are singular, the conventional wisdom that quantum mechanics is a strictly more universal and more accurate theory than classical mechanics is false. But in order to show that a particular singular limit blocks empirical reduction between two theories, it is necessary to explain why empirical reduction between the theories requires this limit to be non-singular in the first place. Berry's discussion does not offer any reason for thinking that the limits he has in mind must be non-singular in order for there to be an accurate quantum description of every real system whose behavior is well-modeled in classical mechanics. The common refrain that classical mechanics should be the limit of quantum mechanics as Planck's constant vanishes seems to reflect a formal rather than an empirical understanding of reduction.



But even if we restrict our interpretation of the above quotation to reduction in a purely formal sense that requires one theory to be a limiting case of another, we are faced with a further ambiguity: namely, that it is not clear what it \textit{means}, in general, for one theory to be a limiting case of another, given that the mathematical concept of a limit is defined only for functions and sets and that a theory is not a well-defined mathematical concept. Given two theories, it is far from clear how, in general, we should determine whether one is a limiting case of another.  Is \textit{any} singular limit relating any two quantities between two theories sufficient to support the conclusion that reduction fails? Or is it only certain special limits that are salient to determining whether reduction occurs? If the latter, on what basis should we determine what these limits are? 

In the next section, I review an argument made by Batterman and following closely on the work of Berry that certain singularities arising in the limit $\hbar \rightarrow 0$ block the reduction of classical to quantum mechanics. I argue that the sense of reduction employed by Batterman is also ambiguous in a number of respects and that we must be careful not to conclude from his analysis that the particular singular limits he discusses pose an obstacle to reduction in the empirical sense.

\subsection{Batterman's Anti-Reductionism Regarding Quantum-Classical Relations}

Like Berry, Batterman concludes that singular limits generally block reduction, and that they do so specifically where the reduction of classical to quantum mechanics is concerned. Batterman takes singular limits to block reduction both in the sense adopted by Berry that requires one theory to be a limiting case of the other, and also in the sense associated with Nagel and Schaffner's well-known account of reduction, which requires that it be possible to derive the laws of one theory from those of another through the use of bridge principles \cite{BattermanDD}.
\footnote{For further discussion of the Nagel/Schaffner approach to reduction, see, for example, \cite{dizadji2010s}.
}
While reduction is understood in Batterman's discussion to fail specifically with respect to the limit-based and Nagel/Schaffner approaches, these approaches themselves are highly vague and therefore open to a wide range of interpretations. 



At the start of his discussion of the relationship between quantum and classical mechanics, Batterman writes, ``The semiclassical limit is singular and no reductive relation obtains between the two theories," where the semiclassical limit  is the limit $\hbar \rightarrow 0$ (see \cite{BattermanDD}, Ch. 7). Recognizing that $\hbar$ is fixed for all real systems and that its numerical value depends on a choice of units, he notes that this limit should be understood as shorthand for the limit $\frac{\hbar}{S_{c}} \rightarrow 0$, where $S_{c}$ is a measure of the ``typical classical action" of the system in question and $\frac{\hbar}{S_{c}}$ is dimensionless (since $\hbar$ also has units of action). However, Batterman's discussion does not specify the appropriate quantitative measure of a system's typical classical action. While this is common practice in discussions of the limit $\hbar \rightarrow 0$, it is important to note that a good deal of physical insight is lost as a result of the failure to give a precise specification of the quantity $S_{cl}$, since knowledge of this quantity would provide a clear delineation of the physical circumstances under which $\hbar$ (understood as shorthand for $\frac{\hbar}{S_{c}}$) can legitimately be regarded as ``small," and under which formulas derived from the assumption of small $\hbar$ can legitimately be applied. 
\footnote{In \cite{landsmanClassicalQuantum}, Landsman offers a number of suggestions as to the particular measure of $S_{cl}$ that might be adopted in various cases.
}
For the moment, though, let us put this worry aside and assume that some appropriate measure of $S_{cl}$ can be be found that physically justifies the assumption of small $\hbar$. 

At the start of his analysis of quantum-classical relations, Batterman writes, ``Given that classical mechanics is supposed to be a limiting case of quantum mechanics as $\hbar \rightarrow 0$, we would like to try to understand the nature of the quantum-mechanical wave functions in this limit." The sense in which this limit is singular is reflected in the breakdown of a particular formal correspondence, due to Maslov, between wavefunctions in Hilbert space and a special class of $N$-dimensional surfaces in $2N$-dimensional classical phase space known as Lagrangian surfaces. Given a generating function $S(q,P)$ of a canonical transformation from coordinates $(q,p)$ to coordinates $(Q,P)$ on classical phase space, the set $\Sigma$ of points in phase space of the form $(q, \nabla_{q}S(q,P))$ for constant $P$ forms a Lagrangian surface. To each such surface, Maslov's method associates the wave function, 

\begin{align} \label{Maslov}
\psi(q) & = K  \bigg| \det \left(\frac{\partial Q_{i}}{\partial q_{j}} \right) \bigg|^{1/2} e^{\frac{i}{\hbar} S(q,P)} \\
& = K  \bigg| \det \left(\frac{\partial^{2} S(q,P)}{\partial q_{j} \partial P_{i}} \right) \bigg|^{1/2} e^{\frac{i}{\hbar} S(q,P)},
\end{align}

\noindent where $Q_{i} = \frac{\partial S(q,P)}{\partial P_{i}}$, $p_{i} = \frac{\partial S(q,P)}{\partial q_{i}}$ and $K$ is an appropriate normalization constant. Batterman notes that the limit $\hbar \rightarrow 0$ of this function is singular in much the same way that the function $\cos(\frac{2 \pi}{\lambda} x)$ is singular as $\lambda \rightarrow 0$: in both cases, the oscillations of the function become infinitely rapid as the relevant parameter approaches zero, so that the limit does not exist. In particular, this singularity blocks the expansion of $e^{\frac{i}{\hbar} S(q,P)}$ in $\hbar$ around the value $\hbar = 0$. 

For fixed, non-zero values of $\hbar$, a Hamiltonian evolution on the classical phase space, which generates an evolution of the Lagrangian surface $\Sigma(t)$, induces a corresponding evolution of the wave function constructed from this surface through Maslov's procedure. In cases where $\Sigma$ develops folds under the Hamiltonian evolution, so that it is multivalued when expressed as a function of the configuration variables $q$, Maslov's construction stipulates that the wavefunction associated with $\Sigma$ is a superposition of different components, one associated with each single-valued ``branch" of $\Sigma$. A special technique devised by Maslov is then used to match these branches along the caustic curves joining them, where $\bigg| \det \left(\frac{\partial^{2} S(q,P)}{\partial q_{j} \partial P_{i}} \right) \bigg|$ is singular. For further details of this construction, see \cite{berry1983semiclassical}.

By contrast with this ``bottom-up" construction, Batterman then considers a ``top-down" derivation of the well-known expression for the WKB wave function, which proceeds by solving the time-independent Schrodinger equation to first order in $\hbar$. The general solution thus derived is,

\begin{align} \label{WKB}
\psi(q) = \frac{A}{\sqrt[4]{2m(E-V(q))}} e^{\frac{i}{\hbar} \int_{x_{0}}^{q} \sqrt{2m(E-V(x))} dx} \ + \ \frac{B}{\sqrt[4]{2m(E-V(q))}} e^{- \frac{i}{\hbar} \int_{x_{0}}^{q} \sqrt{2m(E-V(x))} dx},
\end{align}

\noindent where $A$ and $B$ are arbitrary constants. Batterman observes that at a classical turning point, where $V(x) = E$, the amplitude of the wave function diverges just as the Maslov construction (specifically, the coefficient $\frac{\partial^{2} S(q,P)}{\partial q_{j} \partial P_{i}}$) diverges along the caustics connecting different branches of the Lagrangian surface. 

Given a classical Lagrangian surface $\Sigma$ defined by the function $S(q,P)$ for fixed $P$, one can then fix the constants $A$ and $B$ in the WKB solution at some initial time $t=0$ so as to agree with the wave function (\ref{Maslov}). It is then natural to ask whether the evolution induced on this initial wave function by the classical Hamiltonian evolution of the surface $\Sigma$ approximately agrees with the semi-classical approximation to the Schrodinger evolution associated with (\ref{WKB}). Citing the work of Berry and others, Batterman notes that this agreement holds for a limited class of Lagrangian surfaces - namely, those that are preserved under the classical Hamiltonian evolution. On the other hand, for irregular or chaotic evolutions, Lagrangian surfaces will tend to develop an increasing number of folds - i.e., to become increasingly multivalued in $q$ - over time. Maslov's construction remains applicable in such cases, but only when a certain measure of phase space volume characterizing the separation between adjacent folds in the Lagrangian surface is large in comparison with $\hbar$. The construction becomes inapplicable once the caustics associated with these folds become clustered on scales smaller than $\hbar$. As a result, applying the classical Hamiltonian evolution and then the Maslov construction will not give even approximately the same result as applying the Maslov construction and then the semi-classical Schrodinger dynamics. Where the classical dynamics of the system are chaotic, the semiclassical limit $\hbar \rightarrow 0$ will therefore fail to commute with the infinite time limit $t \rightarrow \infty$ relevant to chaos: for fixed time $t$, one can make $\hbar$ small enough so that Maslov's construction applies and the semiclassical and Hamiltonian evolutions agree; however, if one takes the limit $t \rightarrow \infty$ first, this can't be done. 
\footnote{Another reason often cited for the inability of quantum theory to recover classical chaos is that classical chaos entails exponential divergence of closely spaced initial conditions in phase space, while on the usual association of classical phase space points with narrow wave packet states in quantum theory, the unitary nature of the Schrodinger evolution - which preserves the inner product between any two initial states throughout their evolution - precludes such a divergence between the corresponding wave packets. However, Zurek has argued that when we incorporate the effects of environmental decoherence, the effective quantum dynamics of the system in question is no longer unitary (it is only the total closed system consisting of the system in question \textit{and} its environment that is assumed to evolve unitarily) and so this objection no longer applies. 
}


\subsection{Ambiguities in Batterman's Usage of ``Reduction"}

In his assertion that singular limits block reduction, Batterman clarifies that he understands ``reduction" both in the Nagel/Schaffner (or simply ``Nagelian") sense and in the limit-based sense employed by Berry. However, as I argue here, both of these senses of reduction are subject to a wide range of interpretations, so that a significant degree of ambiguity remains in his  use of the term ``reduction." 

The Nagelian approach remains ambiguous on a range of important issues. In its requirement that it be possible to derive the laws of one theory from those of another through the use of bridge principles, does Nagelian reduction reflect a formal or an empirical sense of reduction? On the one hand, deduction is a formal logical relationship, so that Nagelian reduction could be construed as a kind of formal reduction. On the other hand, one could restrict the requirement of derivability to only those contexts in which the reduced theory successfully describes the behavior of real systems, in which case Nagelian reduction should be regarded as an empirical form of reduction. One can also pose the question: Does Nagelian reduction require a single ``global" derivation of the reduced theory's laws or does it allow for many context-specific ``local" derivations that may employ different bridge principles in different systems? Some treatments of Nagelian reduction require bridge principles to be global, biconditional identity claims while others loosen this requirement to allow for one-way conditional bridge principles that may vary depending on context. There is also a perennial ambiguity as to whether bridge principles are empirically established claims or merely definitions. Batterman's discussion does not make specify which particular understanding of Nagelian reduction he takes to be blocked by singular limits or  \textit{why} singular limits should generally preclude any form Nagelian reduction since nothing inherent to any of these construals relies essentially on taking limits. 

In the case of limit-based reduction, Batterman writes that this sort of reduction rests on the requirement $\lim_{\epsilon \rightarrow 0} T_{f} = T_{c}$, where $T_{f}$ is the more fundamental and $T_{c}$ the less fundamental theory. As discussed above (and as Batterman himself has acknowledged) this requirement is open to a wide range of interpretations. It is unclear what it means to take the limit of a theory or even what is being taken to constitute a theory. We can also pose many of the same questions about limit-based reduction that we posed about Nagelian reduction. Is it being understood formally or empirically? While a formal construal seems more natural, the claim that one theory is a ``limit" or ``limiting case" another is sometimes used loosely to mean simply that the latter supersedes the former, in which case an empirical construal would also be reasonable. One can also ask: is the requirement that one theory be a limit of another local or global? We can imagine a single limit that relates the theories globally, or many context-specific limiting relations that connect the individual models of the theories for different systems. The understanding of limit-based reduction that is hinted at in Batterman's analysis and many other discussions of the limit-based approach seems to be as a global, formal relation, but nothing in Batterman's presentation signals a firm commitment on this point. 
\footnote{For further discussion of the vagueness of existing formulations of the limit-based approach, see \cite{RosalerLocRed}.
}


Where both formal and empirical types of reduction are concerned, it is not made clear in Batterman's analysis why the breakdown of the Maslov construction should be taken to constitute a failure of reduction \textit{per se} - either in the formal or the empirical sense -  rather than merely a failure of one particular \textit{approach} to reduction. While the notion that classical mechanics is the limit of quantum mechanics as $\hbar \rightarrow 0$ comprises what is perhaps the most widespread conventional wisdom about the supposed reduction between quantum and classical mechanics, this does not constitute good reason for thinking that this reduction (in either the formal or the empirical sense) could \textit{only} be effected via the semiclassical limit and Maslov's construction, especially given the vast range of other work on quantum-classical relations that does not invoke this limit or this construction. As I argue in the next section, there are strong reasons to doubt that the technical difficulties associated with Maslov's construction and the singular limits that Batterman highlights pose any obstacle to the empirical reduction of classical to quantum mechanics, and more specifically to attempts to effect this kind of reduction based on decoherence. Given the vagueness of limit-based 
reduction and the existence of other quantum-classical relations that employ the $\hbar \rightarrow 0$ limit in a non-singular way,
\footnote{See Landsman's \cite{landsmanClassicalQuantum} for examples of non-singular $\hbar \rightarrow 0$ limiting relations.
}
further argument is needed to show that difficulties with Maslov's construction block reduction even in a formal sense, and that these difficulties signal anything more than the failure of just one among many attempts to smoothly recover the formalism of classical mechanics from that of quantum mechanics. As it stands, Batterman's argument gives little reason for attributing the sort of broad, sweeping significance to Maslov's construction that it does. 

\subsection{Singular $\hbar \rightarrow 0$ Limits and the Empirical Reduction of Classical to Quantum Mechanics}

It is unclear whether Batterman's analysis is intended to deny the reducibility of classical to quantum mechanics in an empirical sense. However, if he can be fairly interpreted in this way, then it is doubtful that his argument based on singular limits lends much support to this position. Assuming provisionally that Batterman does deny the empirical reducibility of classical to quantum mechanics, his discussion may be criticized on the grounds that it ignores a number of physically salient factors that figure into any realistic quantum description of systems whose behavior is well-modeled by classical mechanics. That is, much of Batterman's analysis focuses primarily on the abstract mathematical formalisms of quantum and classical mechanics, seemingly without a concrete physical system or set of systems in mind. While this sort of methodology may be employed in analyses of formal reduction, greater attention to the features that characterize real systems that we know to behave approximately classically is necessary to address the possibility of empirical reduction. The specific points that Batterman's discussion neglects include the following:

\begin{enumerate}
\item One may make the predictable criticism that Batterman's analysis relies on the limit $\hbar \rightarrow 0$ even though $\hbar$ is constant for all real systems, and that the relevance of the analysis for real systems is obscured by this fact. This sort of criticism may be met, as it is in Batterman's work, with the claim that what is really meant by $\hbar \rightarrow 0$ is $\frac{\hbar}{S_{cl}}  \rightarrow 0$, where the value of $S_{cl}$ \textit{does} vary across real physical systems. Yet, it is more often the case than not - and is the case in Batterman's analysis - that relatively little is said about what the measure $S_{cl}$ should specifically be taken to be. It is also rarely, if ever, the case that expansions in $\hbar$ used to recover classical equations are explicitly re-cast in terms of the dimensionless variable $\frac{\hbar}{S_{cl}}$. While the choice to expand in $\hbar$ rather than $\frac{\hbar}{S_{cl}}$ may be seen as a matter of convenience, it is a convenience that comes at a significant cost to our physical insight, since a specification of the measure $S_{cl}$ is needed to demarcate the circumstances under which $\hbar$ can legitimately be regarded as ``small"; without such a specification, we can only hope that calculations based on the assumption of small $\hbar$ turn out to be physically meaningful in a given context.  
\item Analyses such as those of Batterman, Berry and many others in the field of semiclassical analysis are formulated in the context of isolated quantum systems. But essentially none of the real physical systems whose behavior we know to be well-described by classical mechanics \textit{are}  isolated. Any realistic quantum mechanical model of a classical system such as the center of mass of a golf ball, the moon, or even an alpha particle in a bubble chamber
\footnote{See\cite{JoosZehBook}, Ch. 3,  \cite{bacciaDec} and \cite{barbour2000end}, Ch. 20 for discussion of this last example. 
}
must take account of the fact that these systems are constantly interacting with external degrees of freedom in their environments and thus subject to entanglement with those degrees of freedom. And such entanglement, of course, has significant effects the behavior of the system in question - most notably, the suppression of quantum interference effects.  
However, it should also be pointed out that incorporation of environmental effects, though compelled by the need to give a realistic description of the system in question, substantially complicates matters in certain respects and disrupts the tidy mathematical setting that one finds in the case of isolated systems.
\item Batterman's analysis makes no mention of wave function collapse. However, it is clear that some mechanism for collapse, or effective collapse, must figure into the recovery of real classical behavior from quantum theory. The task of modeling real classical systems in quantum theory does not consist solely in the formal mathematical project of recovering classical \textit{equations} from the formalism of quantum theory, but also in the more conceptual, metaphysical task of understanding how the determinate (or apparently determinate) outcomes characteristic of classical behavior come about in a quantum setting. As with the inclusion of environmental decoherence, the need to recover determinacy from quantum theory brings further complications to the simple mathematical setting of isolated systems evolving unitarily under Schrodinger's equation - in this case, by obliging us to grapple with difficult interpretational issues. In Batterman's analysis, it is not clear when or whether collapses are supposed to occur: is the wave function collapsing continuously, periodically, not at all, or in some other manner? The first of these possibilities would seem to invalidate the WKB approximation that Batterman invokes, which treats the wave function as an approximate plane wave, by requiring the state always to be a narrowly localized wave packet. The second possibility would not successfully recover classical behavior since a position measurement on the sort of plane wave associated with the WKB approximation will give radically unpredictable results for each measurement and so fail to recover classical behavior (since the wave is widely spread out in space). If the wave function never collapses, on the other hand, the question remains as to whether Batterman's analysis is consistent with the appearance of determinate outcomes characteristic of real classical behavior. Without going so far as to demand a solution to the measurement problem, one can still reasonably insist that any realistic attempt to recover classical behavior offer some rough specification as to when collapses or effective collapses occur. 
\end{enumerate}

\noindent Because Batterman's analysis omits a range of crucial, physically salient considerations required for the realistic description of classical systems, it is unlikely that the formal mathematical difficulties that he highlights bear strongly on the empirical reducibility of classical to quantum mechanics. 
\footnote{Here, I take classical behavior to designate behavior that is accurately represented by some \textit{purely} classical model. That is, I do not take it to include those systems whose behavior is well-described by semiclassical models that employ \textit{hybrids} of quantum and classical concepts.   
}
 from quantum theory, which is drawn from the literature on decoherence and does take explicit account of the various factors listed here. This account is spelled out in greater detail in \cite{RosalerIN}, and only its central points and underlying assumptions are discussed here. I argue that this template provides a strategy for the empirical reduction of classical to quantum mechanics that is unaffected by the particular formal mathematical issues that Batterman highlights, for the simple reason that it does not rely on the specific formal construction that Batterman takes to be impeded by singular limits.

\section{An ``Empirical" Approach to Quantum/Classical Reduction} \label{Physical}

Empirical reduction of classical to quantum mechanics requires that every circumstance in which the behavior of some real system is accurately modeled in classical mechanics also is one in which that system's behavior can be modeled at least accurately in quantum mechanics. More precisely, reduction in this sense requires that for every system in the domain of classical mechanics - that is, for every system $S$ whose behavior is accurately characterized by some model of classical mechanics - there exists some model of quantum mechanics, also representing $S$, such that the classical model of $S$ reduces to the quantum model of $S$. Here, reduction between models of a fixed system is taken to require that the reducing model track the system's behavior at least as precisely as the reduced model in all circumstances where the reduced model applies. Thus, empirical reduction between the \textit{theories} of classical and quantum mechanics here is understood to rest on a more basic notion of reduction between two \textit{models} of these theories in cases where both models describe the same, fixed system. Inter-theoretic reduction on this approach may turn out to be a local, piecemeal affair insofar as this approach leaves open the possibility of demonstrating the subsumption of one theory's domain by that of another through numerous system-specific, inter-model reductions. 
(This local, model-based picture of empirical reduction is expounded in greater generality in \cite{RosalerLocRed}.)


The account of empirical reduction of classical to quantum mechanics that I outline here is framed as a template for the reduction between classical and quantum models of a single fixed system in the domain of classical mechanics (e.g., the center of mass of a golf ball, or an alpha particle in a bubble chamber, that traverses an approximately Newtonian trajectory). A complete demonstration of reduction between two models requires proof that certain quantities in the reducing model approximately instantiate the dynamical and other physically salient transformation properties of the reduced model in cases where those features of the reduced model accurately describe the system's behavior. The account of quantum-classical reduction that I give here is formulated as a \textit{template} into which such a proof might be fit, and synthezises a variety of results drawn from the extensive body of work on decoherence theory. This template rests on various assumptions that are conventional in discussions of decoherence and for which proofs have been given in the context of specific models, but which are believed on the grounds of various heuristic arguments to apply generically. My analysis here does not presuppose any specific quantum model, but rather only that the quantum model is compatible with a certain general canonical form of the equation of motion for the reduced density matrix of the system in question.

 
We will see that on this picture of quantum-classical relations, the classical model of the physical system in question is not a special or limiting case of the quantum model; however, the quantum and classical descriptions can be shown to dovetail in certain situations with respect to the behavior of certain variables, and so to provide distinct but mutually consistent (within some margin of approximation) characterizations of the same physical behavior (e.g., the Newtonian trajectory of a golf ball). As an analysis of empirical reduction, this template has certain important advantages over Batterman's: 1) $\hbar$ remains fixed throughout the analysis (as do quantities such as mass and particle number) so that difficulties regarding singular $\hbar \rightarrow 0$ limits are also avoided; 2) the class of quantum models considered here is more realistic as a description of actual classical systems than the models considered in Batterman's discussion, since the models considered here incorporate the effects of the environment on the system's behavior; 3) this account explicitly accommodates the need for a collapse (or effective collapse) mechanism. 



This template starts from the recognition that any realistic quantum mechanical model of a classical system will account for the effects of the system's inevitable interaction with its environment. 
As is typical in models of decoherence, the system of interest, whose classical behavior we wish to recover, and its environment, which includes all degrees of freedom external to the system (including any observer or measuring apparatus that may be present) are modeled together as a closed quantum system with Hilbert space $\mathcal{H}_{S} \otimes \mathcal{H}_{E}$, where $\mathcal{H}_{S}$ is the Hilbert space of the system of interest and $\mathcal{H}_{E}$ the Hilbert space of its environment. The quantum state $| \Psi \rangle$ of the combined system $SE$ is
assumed provisionally to follow the Schrodinger equation, 

\begin{equation} \label{SchrodEnv}
 i \frac{\partial | \Psi \rangle}{\partial t} = \left( \hat{H}_{S} \otimes \hat{I}_{E} + \hat{I}_{S} \otimes \hat{H}_{E} + \hat{H}_{I}  \right) | \Psi \rangle,
\end{equation}

\noindent where $\hat{H}_{S}$ and $\hat{H}_{E}$ operate respectively on states in $\mathcal{H}_{S}$ and $\mathcal{H}_{E}$, $\hat{I}_{E}$ is the identity on $\mathcal{H}_{E}$ , $\hat{I}_{S}$ is the identity on $\mathcal{H}_{S}$,   $\hat{H}_{S} = \frac{\hat{P}_{S}^{2}}{2M_{S}} + V(\hat{X}_{S})$, and $\hat{H}_{I}$, the interaction Hamiltonian, operates on states in $\mathcal{H}_{S} \otimes \mathcal{H}_{E}$. The analysis further assumes that these quantities are such that the reduced density matrix $\hat{\rho}_{S} \equiv Tr_{E}{| \Psi \rangle \langle \Psi |}$ of $S$ follows an equation of the form, 

\begin{equation} \label{DecoherenceMaster}
i \frac{d \hat{\rho}_{S}}{dt} = [ \hat{H}_{S}, \hat{\rho}_{S} ]  - i \Lambda \left[ \hat{X},\left[ \hat{X}  , \hat{\rho}_{S} \right] \right],
\end{equation}

\noindent where $\Lambda$ is a constant that depends on the details of $\hat{H}_{E}$ and $\hat{H}_{I}$, the first term on the right-hand side generates the unitary evolution of  $\hat{\rho}_{S}$, and the second term generates non-unitary effects associated with decoherence, including the suppression of off-diagonal elements of $\langle X' |\hat{\rho}_{S}| X \rangle$.

Zurek \textit{et al}. have argued that there will generically be a separation of timescales over which states in $\mathcal{H}_{S}$ become entangled with $E$ \cite{zurek1993coherent}. Assuming an initial product state for $SE$, most initial pure states of $S$ become entangled with $E$ over the extremely short timescale $\tau_{D}$ associated with decoherence. However, certain special states of $S$, which are known as \textit{pointer states} and which form a basis for $\mathcal{H}_{S}$, suffer entanglement with $E$ only on timescales much longer than the decoherence timescale, and therefore remain as approximate product states for extended time periods. For the specific case where the potential $V$ describes a harmonic oscillator, Zurek \textit{et al} have argued that the pointer states of $S$ should generically be coherent states that are narrowly peaked both in position and in momentum (within the constraints of the uncertainty principle). Further heuristic arguments have been used to show that pointer states should generically be coherent states for more general choices of the potential $V$; see for example Schlosshauer (2008), Ch.Õs 2.8 and 5.2. Assuming this to be the case, environmental decoherence in our class of models lends the overall quantum state evolution of $SE$ a branching structure with respect to a basis of coherent states for $\mathcal{H}_{S}$.  As we will see, this branching can be quantified using the formalism of decoherent histories, originally developed by Gell-Mann, Hartle and Griffiths \cite{gell1993classical}, \cite{griffiths1984consistent}. For a clear discussion of quantum state branching, see  \cite{wallace2012emergent}, Ch. 3.

Given these specifications, our template for the recovery of classical behavior can be summarized as follows. Provisionally assuming a unitary evolution for the closed system consisting of the system $S$ whose classicality we wish to recover and its environment $E$, decoherence lends the total pure state of this system a branching structure. Relative to each branch of the quantum state, the state of the system $S$ is always quasi-classical - that is, always narrowly localized in position and momentum (this is a consequence of the fact that decoherence occurs relative to a pointer basis of coherent states). Thus, one can ascribe a unique quasi-classical trajectory to each branch, given by the evolution of the branch-relative expectation values of $S$'s position and momentum operators. Moreover, a little-discussed form of Ehrenfest's Theorem adapted to open quantum systems entails that on timescales where wave packet spreading - or, more precisely, ensemble wave packet spreading - in $S$ can be neglected, nearly all of these quasi-classical branch trajectories will approximate some solution to Newton's (or Hamilton's) classical equations of motion. Branching still occurs in these cases, but in such a manner that fluctuations associated with branching are confined to small scales of position and momentum. Thus, branch-relative quantum phase space trajectories consist of small, stochastic fluctuations around some deterministic Newtonian trajectory. Each branch of the quantum state is associated with some particular, approximately classical history of the system $S$, corresponding to the sort of localized, approximately Newtonian trajectory that we observe when we see a golf ball moving through the air or the moon orbiting the earth. The manner in which one branch comes to be selected as the ``actual" state of affairs depends on one's interpretation of quantum mechanics and its associated mechanism for effective collapse. However, it should be noted that the significance of the branching structure that emerges through decoherence on this account is that it serves to ``carve out" the set of possible trajectories for the system, from among which some particular interpretation-dependent collapse mechanism serves to select one as the ``outcome."

Let us now outline this story in somewhat more technical detail. Following Wallace (2012), Ch. 3, we can use the pointer states of $S$, which are coherent states $| z \rangle$ - where $z\equiv (q,p)$ denotes the phase space point around which the state $| z \rangle$ is peaked - to construct a positive operator-valued measure (POVM) on $S$'s Hilbert space $\mathcal{H}_{S}$ (see footnote \ref{POVM} for an informal explanation of POVM's). As we will see shortly, this POVM can, in turn, be used to delineate the individual branches of the total quantum state $| \Psi \rangle$. Given an arbitrary partition $\{ \mu_{\alpha} \}$ of the classical phase space $\Gamma_{S}$ of the system $S$, this POVM consists of the set of operators $\{ \hat{\Pi}_{\alpha} \}$ on $\mathcal{H}_{S}$, where the $\hat{\Pi}_{\alpha}$ are defined by

\begin{equation} \label{POVMDef}
\hat{\Pi}_{\alpha} \equiv \int_{\mu_{\alpha}} dz \ |z \rangle \langle z|
\end{equation}

\noindent and $\sum_{\alpha} \hat{\Pi}_{\alpha} = \hat{I}_{S}$, in accordance with the definition of a POVM. Assuming that the coherent states $|z \rangle$ are minimum uncertainty wave packets of $S$, and that the cells $\mu_{\alpha}$ have dimensions in position and momentum that are larger than the position and momentum widths of the state $|z \rangle$ (so that their phase space volume exceeds $\hbar \equiv 1$), the operators in this POVM will also constitute an approximate PVM,
\footnote{\label{POVM}
A projection-valued measure (PVM) on a Hilbert space $\mathcal{H}$ is a set of self-adjoint operators $\{ \hat{P}_{\alpha} \}$ on $\mathcal{H}$ such that

\begin{align}
& \sum_{\alpha} \hat{P}_{\alpha} = \hat{I}, \\
& \hat{P}_{\alpha} \hat{P}_{\beta} = \delta_{\alpha \beta} \hat{P}_{\alpha}, \  \label{PVMOrthog}
\end{align}

\noindent  where there is no summation over repeated indices in (\ref{PVMOrthog}). The concept of a positive operator-valued measure (POVM) on $\mathcal{H}$ generalizes the notion of a PVM by relaxing the requirement of orthogonality in (\ref{PVMOrthog}). Thus, a positive-operator-valued measure (POVM) on a Hilbert space $\mathcal{H}$ is a set $\{ \hat{\Pi}_{\alpha} \}$ of positive operators such that  
 
\begin{align}
& \sum_{\alpha} \hat{\Pi}_{\alpha} = \hat{I}.
\end{align}

\noindent Recall that an operator $\hat{O}$ is positive if it is self-adjoint and $\langle \Psi | \hat{O}|\Psi \rangle \geq 0 $ for every $|\Psi \rangle \in \mathcal{H}$. Note that every PVM is also a POVM. 
}
so that

\begin{equation}
\hat{\Pi}_{\alpha} \hat{\Pi}_{\beta} \approx \delta_{\alpha \beta} \hat{\Pi}_{\alpha}.
\end{equation}

\noindent We can then extend the set $\{ \hat{\Pi}_{\alpha} \}$ to an approximate PVM on the total Hilbert space $\mathcal{H}_{S} \otimes \mathcal{H}_{E}$ of the system $SE$ by defining the set of operators $\{ \hat{P}_{\alpha} \}$, where $\hat{P}_{\alpha} \equiv \hat{\Pi}_{\alpha} \otimes \hat{I}_{E}$, with $\hat{I}_{E}$ the identity on $\mathcal{H}_{E}$. In what follows, we will make use of this approximate PVM to analyze the branching structure of the pure state evolution of the closed system $SE$. This particular choice of approximate PVM is motivated by the fact that the coherent states $|z \rangle$ are pointer states of $S$ under its interaction with $E$; this entails that the different branch states defined using this PVM will be mutually orthogonal at each time, which in turn entails the existence of a branching structure for the overall state evolution. 

To see this more clearly, let us decompose the unitary evolution of the pure state of $SE$ using the approximate coherent state PVM just defined. Dividing the time interval between time $0$ and time $t > > \tau_{D}$ into $N$ equal steps $\Delta t = \frac{t}{N}$, we can represent the unitary state evolution as follows:

\begin{align} \label{Histories}
| \Psi( t)\rangle &=  e^{- i\hat{H} N \Delta t} | \Psi_{0} \rangle \\
& =  \sum_{\alpha_{0},...,\alpha_{N}} \hat{C}_{\alpha_{0}...\alpha_{N}} |  \Psi_{0}  \rangle, 
\end{align} 

\noindent where 
$ \hat{C}_{\alpha_{0}...\alpha_{N}} |  \Psi_{0}  \rangle \equiv \hat{P}_{\alpha_{N}}  e^{-\frac{i}{\hbar}\hat{H} \Delta t}\hat{P}_{\alpha_{N-1}} ...  e^{-i \hat{H} \Delta t} \hat{P}_{\alpha_{1}} e^{-\frac{i}{\hbar}\hat{H} \Delta t} \hat{P}_{\alpha_{0}}   |  \Psi_{0}  \rangle$. It can be seen straightforwardly from the definition of the POVM operators in (\ref{POVMDef}) that environmental decoherence relative to a pointer basis of coherent states for $S$ entails (by virtue of the definition of the operators $\hat{\Pi}_{\alpha}$) that at each time $N \Delta t$, these branch vectors satisfy the condition

\begin{equation} \label{HistoriesDecoherence}
\langle \Psi_{0} | \hat{C}_{\alpha'_{0} \alpha'_{1} ... \alpha'_{N}}^{ \dagger} \hat{C}_{\alpha_{0} \alpha_{1}...\alpha_{N}}   | \Psi_{0}\rangle \approx 0 
\end{equation}

\noindent for $\alpha_{k} \neq \alpha'_{k}$ for any $0 \leq k \leq N$. Thus, the unitary evolution of the total quantum state follows the progression,

\

\noindent \underbar{Unitary Branching Evolution}
\begin{flalign} \label{BranchSequence}
| \Psi_{0} \rangle \xrightarrow{ e^{-i \hat{H} \Delta t}}    \sum_{\alpha_{1}} \hat{C}_{\alpha_{1}} | \Psi_{0} \rangle \xrightarrow{ e^{-i \hat{H} \Delta t}}   \sum_{\alpha_{1} \alpha_{2}} \hat{C}_{\alpha_{1} \alpha_{2}} | \Psi_{0} \rangle \xrightarrow{ e^{-i \hat{H} \Delta t}} ... \xrightarrow{ e^{-i \hat{H} \Delta t}}  \sum_{\alpha_{1} \alpha_{2} ... \alpha_{N}} \hat{C}_{\alpha_{1} \alpha_{2} ... \alpha_{N}} | \Psi_{0} \rangle,   
\end{flalign}

\noindent where for every $1\leq i \leq N$, $\langle \Psi_{0} | \hat{C}_{\alpha'_{0} \alpha'_{1} ... \alpha'_{i}}^{ \dagger} \hat{C}_{\alpha_{0} \alpha_{1}...\alpha_{i}}   | \Psi_{0}\rangle \approx 0$ if $\alpha_{k} \neq \alpha'_{k}$ for any $0 \leq k \leq i$. Each index added to the sum at each new time step $\Delta t$ captures a separate branching of the quantum state. 

Thus far, our analysis has assumed that the quantum state of the system $SE$ evolves unitarily according to Schrodinger's equation, (\ref{SchrodEnv}). However, the world of our experience - which is characterized by determinate values of position and momentum for systems like the ones we see behaving classically - can only be associated with one particular branch. Given the branching structure that is carved out by the unitary dynamics through decoherence, we can associate to each branch at time $t= i\Delta t$ an effective, normalized ``branch state," 
  $\frac{1}{W_{\alpha_{1}...\alpha_{i}}}\hat{C}_{\alpha_{1} ... \alpha_{i}} | \Psi_{0} \rangle$, where $W_{\alpha_{1}...\alpha_{i}} \equiv  \sqrt{ \langle \Psi_{0}|\hat{C}^{\dagger}_{\alpha_{1} ...\alpha_{i}}   \hat{C}_{\alpha_{1} ...\alpha_{i}} | \Psi_{0} \rangle} $. It is then possible to define an effective, stochastic evolution for the branch-relative state of the system $SE$ as follows:

\

\noindent \underbar{Stochastic Branch-Relative State Evolution} \label{BRStateEv}

\begin{flalign} \label{BranchSequence}
& \frac{1}{W_{\alpha_{1}}} \hat{C}_{\alpha_{1}} | \Psi_{0} \rangle \xrightarrow{prob. \frac{|W_{\alpha_{1} \alpha_{2}}|^{2}}{|W_{\alpha_{1}}|^{2}}}    \frac{1}{W_{\alpha_{1} \alpha_{2}} } \hat{C}_{\alpha_{1} \alpha_{2}} | \Psi_{0} \rangle \xrightarrow{prob. \frac{|W_{\alpha_{1} \alpha_{2} \alpha_{3}}|^{2}}{|W_{\alpha_{1} \alpha_{2}}|^{2}}}    \frac{1}{W_{\alpha_{1} \alpha_{2} \alpha_{3}} } \hat{C}_{\alpha_{1} \alpha_{2} \alpha_{3}} | \Psi_{0} \rangle  \xrightarrow{prob. \frac{|W_{\alpha_{1} \alpha_{2} \alpha_{3} \alpha_{4}}|^{2}}{|W_{\alpha_{1} \alpha_{2} \alpha_{3}}|^{2}}} \\ 
& \  \nonumber \\
&\ \ ... \ \ \xrightarrow{prob. \frac{|W_{\alpha_{1} \alpha_{2} ... \alpha_{N-1} \alpha_{N}}|^{2}}{|W_{\alpha_{1} \alpha_{2} ... \alpha_{N-1}}|^{2}}}  \frac{1}{W_{\alpha_{1} \alpha_{2}...\alpha_{N}}}  \hat{C}_{\alpha_{1} \alpha_{2} ... \alpha_{N}} | \Psi_{0} \rangle. \nonumber
\end{flalign} 

\noindent It is straightforward to see that the transition probability $\bigg| \frac{W_{\alpha_{1}...\alpha_{i} \alpha_{i+1}}}{W_{\alpha_{1}...\alpha_{i}}}\bigg|^{2}$ at the $i^{th}$ time step is simply the square magnitude of the $\alpha_{i+1}^{th}$ component of the time evolved state $e^{-i \hat{H} \Delta t}\frac{1}{W_{\alpha_{1}...\alpha_{i}}}\hat{C}_{\alpha_{1} ... \alpha_{i}} | \Psi_{0} \rangle =  \frac{1}{W_{\alpha_{1}...\alpha_{i}}} \sum_{\alpha_{i+1}} \hat{C}_{\alpha'_{1} ... \alpha'_{i} \alpha_{i+1}} | \Psi_{0} \rangle $, in accordance with the usual Born Rule prescription for collapse. 
The physical justification for taking a single branch state rather than the overall superposition as the effective state of $SE$ and for adopting the above succession of Born Rule collapses as the effective evolution of this state will depend on the particular collapse mechanism and interpretation of quantum mechanics that one takes as the basis of the analysis. In \cite{RosalerIN}, I discuss the manner in which various realist interpretations purport to underwrite this sequence of collapses or effective collapses, and in \cite{RosalerDBB2015} show in more detail how this interpretation-neutral  account can be specially tailored to the specific context of the de Broglie-Bohm interpretation. 

My central concern is with questions of reduction and not with questions concerning the interpretation of quantum mechanics. 
Putting questions of interpretation to the side and assuming that our effective, stochastic, branch-relative state evolution above is underwritten by \textit{some} mechanism for collapse or effective collapse, we can associate a point in classical phase space $\Gamma_{S}$ to each effective branch state at each time by taking the branch-relative expectation values of SÕs position and momentum operators:

\begin{empheq}[box=\fbox]{align}
 X^{q}_{\alpha_{1}...\alpha_{N}}(N \Delta t) & \equiv \frac{1}{|W_{\alpha_{1}...\alpha_{N}}|^{2}} \langle \Psi_{0} |    \hat{C}^{\dagger}_{\alpha_{1}...\alpha_{N}} \left( \hat{X}_{S} \otimes \hat{I}_{E} \right)  \hat{C}_{\alpha_{1}...\alpha_{N}} | \Psi_{0} \rangle \\
  & \ \ \\
 P^{q}_{\alpha_{1}...\alpha_{N}}(N \Delta t) & \equiv \frac{1}{|W_{\alpha_{1}...\alpha_{N}}|^{2}} \langle \Psi_{0} |    \hat{C}^{\dagger}_{\alpha_{1}...\alpha_{N}} \left( \hat{P}_{S} \otimes \hat{I}_{E} \right)  \hat{C}_{\alpha_{1}...\alpha_{N}} | \Psi_{0} \rangle.
\end{empheq}

\noindent Furthermore, to each stochastic, branch-relative state evolution (\ref{BranchSequence}) we may associate the following stochastic, quasi-classical  evolution in classical phase space:

{\footnotesize
\begin{flalign*}
& \left( X^{q}_{\alpha_{1}}\left(\Delta t \right), P^{q}_{\alpha_{1}}( \Delta t) \right)   \xrightarrow{prob. \frac{|W_{\alpha_{1} \alpha_{2}}|^{2}}{|W_{\alpha_{1}}|^{2}}}   
\left( X^{q}_{\alpha_{1} \alpha_{2}}(2 \Delta t), P^{q}_{\alpha_{1} \alpha_{2}}\left(2 \Delta t \right) \right)      \xrightarrow{prob. \frac{|W_{\alpha_{1} \alpha_{2} \alpha_{3}}|^{2}}{|W_{\alpha_{1} \alpha_{2}}|^{2}}}   \left( X^{q}_{\alpha_{1} \alpha_{2} \alpha_{3}}\left(3 \Delta t\right), P^{q}_{\alpha_{1} \alpha_{2} \alpha_{3}}\left(3 \Delta t \right) \right) \\
& \xrightarrow{prob. \frac{|W_{\alpha_{1} \alpha_{2} \alpha_{3} \alpha_{4}}|^{2}}{|W_{\alpha_{1} \alpha_{2} \alpha_{3}}|^{2}}}  \ \ ... \ \ \xrightarrow{prob. \frac{|W_{\alpha_{1} \alpha_{2} ... \alpha_{N-1} \alpha_{N}}|^{2}}{|W_{\alpha_{1} \alpha_{2} ... \alpha_{N-1}}|^{2}}}  \left( X^{q}_{\alpha_{1} \alpha_{2} \alpha_{3}...\alpha_{N}}(N \Delta t), P^{q}_{\alpha_{1} \alpha_{2} \alpha_{3}...\alpha_{N}}(N \Delta t) \right). 
\end{flalign*} 
}

\noindent Note that the discreteness of the timesteps $\Delta t$ here is superimposed artificially on a dynamical process of entanglement and branching that occurs continuously in time. By making the time intervals $\Delta t$ successively smaller, our discrete, stochastic, branch-relative phase space trajectories provide successively better approximations to a continuous, stochastic ``quantum" phase space trajectory $\left(X^{q}\left(t\right), P^{q}\left(t\right)\right)$ for the system $S$.

Thus far, my analysis has suggested how stochastic quasi-classical trajectories can be recovered from our effective, branch-relative state evolution. However, I have not given an argument as to why these quasi-classical trajectories should approximate solutions to classical equations of motion (e.g., Newton's or Hamilton's equations). Taking $\left(X^{q}\left(t\right), P^{q}\left(t\right)\right)$ to represent the phase space trajectory associated with the stochastic evolution of the branch-relative expectation values of position and momentum, and $\left(X^{c}\left(t\right), P^{c}\left(t\right)\right)$ to represent the deterministic classical trajectory with the same initial condition as this quantum trajectory, so that $X^{q}(0) = X^{c}(0)$ and $P^{q}(0) = P^{c}(0)$, the recovery of classical behavior requires us to show that, with probability $1- \epsilon$, for very small $\epsilon$, the relation

\begin{empheq}[box=\fbox]{align} \label{QCApprox}
& \bigg| X^{q}( t) - X^{c}( t)  \bigg| < 2\delta_{X} \\
& \bigg| P^{q}( t) - P^{c}(t)  \bigg| < 2 \delta_{P} \nonumber,
\end{empheq}

\noindent holds for $ t < \tau$, where $\delta_{X}$ and $\delta_{P}$ are the margins of error and $\tau$ is the timescale within which the Newtonian model is known to approximate the behavior of the system $S$ itself. The factor of $2$ on the right hand side of (\ref{QCApprox}) has been included because, if both the quantum and classical models track the system's behavior within margin $\delta_{Z} \equiv \left(\delta_{X}, \delta_{P} \right)$, they can differ at most by $2 \delta_{Z}$. A seldom-discussed form of Ehrenfest's Theorem, formulated for a certain class of open systems $S$ (the familiar form of Ehrenfest's Theorem only concerns closed systems), ensures that 

\begin{equation}
\frac{d \ Tr[\hat{\rho}_{S}  \hat{P}] }{dt} =  -  Tr[ \hat{\rho}_{S} \hat{\frac{\partial V(X) }{ \partial X}}],
\end{equation}

\noindent or more concisely,

\begin{equation} \label{OpenEhrenfest}
\frac{d \langle \hat{P} \rangle}{dt} =  - \langle \hat{\frac{\partial V(X) }{ \partial X}} \rangle,
\end{equation}

\noindent where $\hat{\rho}_{S}$ is the reduced density matrix of $S$, $\langle \hat{P} \rangle = Tr [{\hat{\rho}_{S} \hat{P}}] $ and $\langle \hat{X} \rangle = Tr [{\hat{\rho}_{S} \hat{X}}] $ \cite{JoosZehBook}. If we now impose the restriction to density matrices $\hat{\rho}_{S}$ for which the position-space probability distribution $ \langle X | \hat{\rho}_{S}| X \rangle$ is narrowly peaked about some particular $X_{0}$, with a width that is small by comparison with the characteristic length scales of $V$, it follows that

\begin{equation} \label{OpenEhrenfestApprox}
\frac{d \langle \hat{P} \rangle}{dt} \approx  - \frac{\partial V(X) }{ \partial X}\bigg|_{\langle \hat{X} \rangle}.
\end{equation}

\noindent Combined with the relation $\frac{d \langle \hat{X} \rangle }{dt} = \frac{\langle  \hat{P} \rangle}{M}$, this entails that the expectation values $\langle \hat{X} \rangle$ and $\langle \hat{P} \rangle$ follow an approximately Newtonian trajectory as long as $ \langle X | \hat{\rho}_{S}| X \rangle$ remains suitably narrow by comparison with the dimensions of $V$, and to a measure of approximation determined by the width of the distribution $ \langle X | \hat{\rho}_{S}| X \rangle$. Thus, the timescales on which the expectation values $\langle \hat{X} \rangle$ and $\langle \hat{P} \rangle$ follow an approximately Newtonian trajectory will depend on the rate at which the ``ensemble" distribution $ \langle X | \hat{\rho}_{S}| X \rangle$ spreads over time. 
\footnote{It is important to note that the ensemble distribution $ \langle X | \hat{\rho}_{S}(t) | X \rangle$ reflects a distribution \textit{across} branches, which originate in some initial narrow state $\hat{\rho}^{0}_{S}$.
} 
As I argue in further detail in \cite{RosalerIN}, it follows that over timescales where ensemble spreading can be ignored, the only branch-relative phase space trajectories that have non-negligible probability of occurring on our stochastic, branch-relative phase space evolution are ones that are approximately (i.e., to within margin of error given by the width of the distribution $\langle X | \hat{\rho}_{S}(t) | X \rangle$) Newtonian in form. As a general rule of thumb, these timescales will be larger when the mass of $S$ is large, and smaller when the effects of chaos - as quantified, for example, by the Lyapunov exponent in the classical Hamiltonian $H_{S}$ -  are significant. Environmental decoherence also contributes somewhat to the rate of ensemble (as opposed to coherent) spreading; see \cite{schlosshauer2008decoherence}, Ch. 3 for discussion of this point. 

To complete this analysis, one must show that the timescales on which ensemble spreading in $S$ can be ignored are at least as long as the timescales over which Newtonian trajectories are known to approximate of the actual trajectory of the system in question. Otherwise, the branch-relative phase space evolutions prescribed by the quantum model of the system would not describe the behavior of the system at least as well as the classical model, and empirical reduction of the classical to the quantum model would fail. Calculation of the timescales for ensemble spreading requires more detailed specification of the particular quantum model of the the system in question. The timescales and margins of error within which classical Newtonian trajectories can be expected to track the system's behavior are a matter for empirical investigation; from a practical point of view, it may only be possible to specify them imprecisely within certain broadly specified bounds. Moreover, as was already discussed, one must provide some justification for the \textit{ad hoc} collapse posited by (\ref{BranchSequence}). Competing interpretations of quantum theory all offer different accounts of collapse. For further discussion of the manner in which the collapse mechanisms associated with different interpretations can be fit into this account of classical behavior, see \cite{RosalerIN}. 

On this account of empirical reduction between quantum and classical models of the same system,  the quantum mechanical quantities $\frac{1}{|W_{\alpha_{1}...\alpha_{N}}|^{2}} \langle \Psi_{0} |    \hat{C}^{\dagger}_{\alpha_{1}...\alpha_{N}} \left( \hat{X}_{S} \otimes \hat{I}_{E} \right)  \hat{C}_{\alpha_{1}...\alpha_{N}} | \Psi_{0} \rangle$ and $\frac{1}{|W_{\alpha_{1}...\alpha_{N}}|^{2}} \langle \Psi_{0} |    \hat{C}^{\dagger}_{\alpha_{1}...\alpha_{N}} \left( \hat{P}_{S} \otimes \hat{I}_{E} \right)  \hat{C}_{\alpha_{1}...\alpha_{N}} | \Psi_{0} \rangle$ represent the very same features of the system that are represented by the phase space point $\left(X^{c}\left(t\right), P^{c}\left(t\right)\right)$ in the classical model of $S$. In this sense, the two variables - quantum and classical - may said to co-refer, even if their agreement is only approximate.  If the assumptions of the above analysis are correct, we see that whatever can be modeled classically in terms of a point evolving in classical phase space can be modeled quantum mechanically in terms of the effective, stochastic quantum evolution of branch-relative expectation values of the system's position and momentum operators. 


The above analysis supports something close to Belot's critique of Batterman's claim that quantum mechanics is explanatorily inadequate because it requires reference to the resources of classical mechanics for the explanation of certain phenomena. Belot writes that ``the mathematics of the less fundamental theory is definable in terms of the mathematics of the fundamental theory and ... only the latter need be given a physical interpretation ... so we can view the desired explanation as drawing only on the resources internal to the more fundamental theory" \cite{belot2005whose}. While it is not quite correct on our analysis to say that a classical phase space point is \textit{defined} in terms of the branch-relative expectation values of position and momentum, we can say that both quantities represent the same physical degrees of freedom, so that any physical system whose behavior is well-described by the classical evolution of a point in phase space can equally well be described by the quantum-mechanical evolution of branch-relative expectation values of position and momentum. Thus, quantum mechanics does not require classical mechanics for the description even of phenomena that lie properly within the domain of classical mechanics and offers its own counterparts to any physically salient features of classical theory.


The analysis of quantum-classical relations given in this section provides a general strategy for understanding how any situation that is successfully modeled in classical mechanics can also be modeled at least as precisely in quantum mechanics, and so provides the outline of an account of the empirical reduction of classical mechanics to quantum mechanics. It enables us to understand the behavior of any real classical system - say, the center of mass of a golf ball in mid-flight - not only in the familiar setting of classical mechanics, but also in the more intricate framework of quantum mechanics.

\subsection{Open Questions Concerning the Decoherence-Based Account of Empirical Quantum-Classical Reduction}



It is doubtful that the singular limits highlighted in Batterman's analysis have much, if any, bearing on the viability of the decoherence-based approach to empirical reduction that I have outlined here. The basic assumptions, methodologies and concerns guiding the two analyses differ too widely. However, it is still reasonable to have doubts about the sort of decoherence-based account of empirical reduction described above insofar as it relies heavily on a number of strong and unproven assumptions. Among the most important of these unproven assumptions are the following: 



\begin{enumerate}
\item It was assumed that the pointer states of the system $S$ are narrowly peaked wave packets. Zurek and others have argued that this is a fairly generic feature of systems of the sort that have concerned us here. However, such arguments have a strongly heuristic character, and it is important to ask whether it is possible to give a more rigorous justification of the claim that narrowly peaked coherent states are generically pointer states of the systems we know to behave classically. 
\item It was assumed implicitly in our analysis that the branching of the quantum state associated with decoherence is an effectively irreversible process. By close analogy with the problem of the arrow of time in classical statistical mechanics, this irreversibility must be reconciled with the time-reversal symmetry of the fundamental Schrodinger evolution. The question of how or whether this reconciliation can be effected remains open. For one account of the origins of irreversibility associated with branching of the quantum state, see \cite{wallace2012emergent}, Ch. 9.
\item The open-systems version of Ehrenfest's Theorem, which was used to show that branch-relative phase space trajectories are approximately Newtonian on timescales where ensemble spreading is small, assumes that the evolution of the reduced density matrix $\hat{\rho}_{S}$ of the system $S$ is governed by a "master" equation of the specific form given above (see \cite{JoosZehBook}). A complete reduction of classical to quantum here requires a derivation of this master equation from the microscopic Schrodinger equation for the closed combination of system and environment. It is important to note, however, that the details of the microscopic quantum model may vary significantly from one classical system to another. 
\item It should be shown explicitly for the system in question that the rate of ensemble wave packet spreading in the quantum model is consistent with the timescales on which solutions to classical equations of motion are known to track the system's behavior. As was shown above, faster ensemble spreading causes the total quantum state to branch more rapidly, which in turn makes it more likely that branch-relative trajectories will exhibit significant stochastic fluctuations away from classicality on shorter timescales. One must show that ensemble spreading is sufficiently slow so that with probability extremely close to $1$, the branch-relative, quantum phase space trajectory of the system $S$ will approximate the classical trajectory with the same classical initial conditions over timescales for which the classical trajectory tracks the system's behavior. In answer to concerns about the capacity of quantum theory to recover classically chaotic behavior, Zurek has argued that decoherence addresses these worries. However, this is by no means a consensus view. For discussion of the connection between decoherence and chaos, see \cite{zurek1995quantum}, \cite{zurek1998decoherence}, and \cite{bokulich2008reexamining}, Ch.1, including references therein. 
\end{enumerate} 

\noindent Because of these remaining open questions, the decoherence-based framework for recovering classical behavior outlined above should be viewed as a research program that consists partly in providing further support for its underlying assumptions, and also in extending it to new cases.

\section{Conclusion} \label{Conclusion}

I have juxtaposed two alternative analyses of the purported reduction of classical to quantum mechanics. The first of these analyses, due to Batterman, denies the possibility of reducing classical to quantum mechanics on the grounds that the limit $\hbar \rightarrow 0$ is singular and that a certain formal construction associating wave functions with a special class of surfaces in classical phase space breaks down in chaotic systems. The second analysis, drawn from the literature on decoherence theory, suggests that it is possible to effect a reduction of classical to quantum mechanics by incorporating the system's interaction with its environment and paying careful attention to the branching evolution for the quantum state that results. By distinguishing between ``formal" and ``empirical" senses of reduction, I have argued the decoherence-based account offers a viable strategy for empirical reduction while the issues raised by Batterman potentially pose problems for reduction only in the formal sense. 

As Berry has noted, questions about inter-theoretic reduction are highly mathematical in nature. However, if our concern is with the type of reduction that ensures each successive theory strictly includes  the domain of applicability of its predecessor - as is often supposed to be the case between classical and quantum mechanics - then it is a mistake to think that reduction in this sense is \textit{solely} a question of the mathematical relationship between two theories. Empirical facts about the domain of success of the reduced theory must also be taken into account, and it is the empirical rather than the formal sense of reduction that is relevant. Questions about empirical reduction are strongly mathematical in nature, but also rely on empirical determinations of the quality and scope of agreement between the reduced model and the system itself. Therefore, reduction in this sense is not a two-place relation between theories but, in some sense, a three-place relation connecting the theories \textit{and} the real physical systems they describe. It is likely that the distinction between formal and empirical reduction, and between formal and empirical approaches to inter-theory relations more generally, can usefully be applied to the study of other inter-theory relations as well. 

\

\

\noindent \textbf{Acknowledgments:}
Thanks to David Wallace, Simon Saunders, Christopher Timpson, and Jeremy Butterfield for many helpful discussions on the classical domain of quantum theory.

\def\bibsection{\section*{References}}
\scriptsize
\bibliographystyle{plain}
\bibliography{DPhilRefs}

\begin{thebibliography}{10}

\bibitem{bacciaDec}
Guido Bacciagaluppi.
\newblock The role of decoherence in quantum mechanics.
\newblock In Edward~N. Zalta, editor, {\em The Stanford Encyclopedia of
  Philosophy}. Winter 2012 edition, 2012.

\bibitem{barbour2000end}
Julian Barbour.
\newblock {\em The end of time: The next revolution in physics}.
\newblock Oxford University Press, 2000.

\bibitem{BattermanDD}
R.~Batterman.
\newblock {\em The Devil in the Details: Asymptotic Reasoning in Explanation,
  Reduction, and Emergence}.
\newblock Oxford University Press, 2002.

\bibitem{belot2005whose}
Gordon Belot.
\newblock Whose devil? which details?*.
\newblock {\em Philosophy of Science}, 72(1):128--153, 2005.

\bibitem{BerryAsymptotics}
M.~Berry.
\newblock Asymptotics, singularities and the reduction of theories.
\newblock In Brian~Skyrms Dag~Prawitz and Dag Westerst{\aa}hl, editors, {\em
  Logic, Methodology and Philosophy of Science, IX: Proceedings of the Ninth
  International Congress of Logic, Methodology and Philosophy of Science,
  Uppsala, Sweden, August 7--14, 1991 (Studies in Logic and Foundations of
  Mathematics: Volume 134)}, volume 134, pages 597--607, 1994.

\bibitem{berry1983semiclassical}
MV~Berry.
\newblock Semiclassical mechanics of regular and irregular motion.
\newblock {\em Les Houches lecture series}, 36:171--271, 1983.

\bibitem{bokulich2008reexamining}
Alisa Bokulich.
\newblock {\em Reexamining the Quantum-Classical Relation}.
\newblock Cambridge University Press Cambridge, 2008.

\bibitem{dizadji2010s}
F.~Dizadji-Bahmani, R.~Frigg, and S.~Hartmann.
\newblock Who's afraid of {N}agelian reduction?
\newblock {\em Erkenntnis}, 73(3):393--412, 2010.

\bibitem{gell1993classical}
M.~Gell-Mann and J.B. Hartle.
\newblock Classical equations for quantum systems.
\newblock {\em Physical Review D}, 47(8):3345, 1993.

\bibitem{griffiths1984consistent}
R.B. Griffiths.
\newblock Consistent histories and the interpretation of quantum mechanics.
\newblock {\em Journal of Statistical Physics}, 36(1):219--272, 1984.

\bibitem{JoosZehBook}
E.~Joos, D.~Zeh, C.~Kiefer, D.~Giulini, J.~Kupsch, and I.-O. Stamatescu.
\newblock {\em Decoherence and the Appearance of a Classical World in Quantum
  Theory}.
\newblock Springer, second edition edition, 2003.

\bibitem{landsmanClassicalQuantum}
N.P. Landsman.
\newblock Between classical and quantum.
\newblock In J.~Butterfield and J.~Earman, editors, {\em Philosophy of Physics
  (Handbook of the Philosophy of Science)}, volume~1. Elsevier, 2007.

\bibitem{RosalerLocRed}
Joshua Rosaler.
\newblock Local reduction in physics.
\newblock {\em Studies in History and Philosophy of Modern Physics}, 50:54--69,
  2015.

\bibitem{RosalerIN}
Joshua Rosaler.
\newblock Interpretation neutrality in the classical domain of quantum theory.
\newblock {\em Studies in History and Philosophy of Modern Physics},
  forthcoming, 2015.

\bibitem{RosalerDBB2015}
Joshua Rosaler.
\newblock Is de broglie-bohm theory specially equipped to recover classical
  behavior?
\newblock {\em Philosophy of Science}, forthcoming, 2015.

\bibitem{sakurai1995modern}
J.J. Sakurai, S.F. Tuan, and E.D. Commins.
\newblock Modern quantum mechanics.
\newblock {\em American Journal of Physics}, 63:93, 1995.

\bibitem{schlosshauer2008decoherence}
M.A. Schlosshauer.
\newblock {\em Decoherence and the Quantum-To-Classical Transition}.
\newblock Springer, 2008.

\bibitem{wallace2012emergent}
D.~Wallace.
\newblock {\em The Emergent Multiverse: Quantum Theory According to the Everett
  Interpretation}.
\newblock Oxford University Press, Oxford, 2012.

\bibitem{zurek1993coherent}
W.H. Zurek, S.~Habib, and J.P. Paz.
\newblock Coherent states via decoherence.
\newblock {\em Physical Review Letters}, 70(9):1187--1190, 1993.

\bibitem{zurek1995quantum}
W.H. Zurek and J.P. Paz.
\newblock Quantum chaos: a decoherent definition.
\newblock {\em Physica D: Nonlinear Phenomena}, 83(1):300--308, 1995.

\bibitem{zurek1998decoherence}
Wojciech~H Zurek.
\newblock Decoherence, chaos, quantum-classical correspondence, and the
  algorithmic arrow of time.
\newblock {\em Physica Scripta}, 1998(T76):186, 1998.

\end{thebibliography}

\end{document}